\documentclass{sig-alternate-nocopyright}

\usepackage{booktabs} 
\usepackage{times}

\usepackage{url,bm}
\usepackage{latexsym}

\usepackage{algpseudocode} 
\usepackage{multirow}

\usepackage{etoolbox}
\newtoggle{conf} \toggletrue{conf}

\newcommand{\cut}[1]{}

\begin{document}
\title{Dionysius: A Framework for Modeling Hierarchical User\\ Interactions in Recommender Systems}

\numberofauthors{1}
\author{\alignauthor 
Jian Wang, \ Krishnaram Kenthapadi, \ Kaushik Rangadurai, \ David Hardtke\thanks{Work done while all authors were affiliated with LinkedIn. Current affiliation of the following authors: Jian Wang and Kaushik Rangadurai (Google); David Hardtke (Pandora).}\\
\affaddr{LinkedIn Corporation, USA}
}

\maketitle

\begin{abstract}
Real-time large-scale personalized recommendation systems power several user-facing products at many social media and web platforms. To meet business requirements, such applications must score millions of structured candidate items for each query, offer a high degree of data freshness, and respond with low latency. To address these challenges, many such systems make use of content-based recommendation models based on logistic regression. A fundamental problem with content-based models is that they are based primarily on the explicit user context in the form of user profile, but do not take into account implicit user context in the form of user interactions.

We address the following problem: {\em How do we incorporate user item interaction signals as part of the relevance model in a large-scale personalized recommendation system such that, (1) the ability to interpret the model and explain recommendations is retained, and (2) the existing infrastructure designed for the (user profile) content-based model can be leveraged?} We propose {\em Dionysius}, a hierarchical graphical model based framework and system for incorporating user interactions into recommender systems, with minimal change to the underlying infrastructure. We learn a hidden fields vector for each user by considering the hierarchy of interaction signals, and replace the user profile-based vector with this learned vector, thereby not expanding the feature space at all. Thus, our framework allows the use of existing recommendation infrastructure that supports content based features. We implemented and deployed this system as part of the recommendation platform at LinkedIn for more than one year. We validated the efficacy of our approach through extensive offline experiments with different model choices, as well as online A/B testing experiments. Our deployment of this system as part of the job recommendation engine resulted in significant improvement in the quality of retrieved results, thereby generating improved user experience and positive impact for millions of users.
\end{abstract}

\section{Introduction}
\label{SecIntro}

Personalized recommendation systems form the backbone of several user-facing products at many large Internet companies. Such systems pose unique challenges, wherein personalized recommendations need to be computed in real-time from millions of structured candidate items while providing a high degree of data freshness. To address these challenges, many such systems in practice make use of content-based recommendation models based on logistic regression. Content-based models can be implemented by incorporating content related features across structured user fields and/or structured item fields as part of the model, in which the feature space is usually static and limited. To use the job recommendation domain as an example, structured user fields could include the user's current job title, seniority, skills, working experience, education experience, and so on. Structured job fields could include the job title, function, industry, company, seniority, skills, description, and so on. Content related features could be defined in terms of similarity between different combinations of user fields and job fields. Such features are easy to compute in both the offline training pipeline and the online prediction stage. Another benefit is that the resulting model can be easily interpreted, and explanations could be provided for recommendations. 

A fundamental problem with content-based models is that they are based primarily on the explicit user context (say, in the form of user profile), but do not take into account the implicit user context in the form of user interactions. However, two users with near identical profiles may have very different preferences for job recommendations: the first user may prefer jobs with similar titles as her current position from a few selected companies, while the second user may prefer jobs with identical title as her current position but from a wider range of companies. 

In this paper, we address the problem of incorporating user item interaction signals as part of the relevance model in a large-scale personalized recommendation system, while retaining the ability to interpret the model and leveraging existing recommendation system infrastructure.
We propose {\em Dionysius}, a hierarchical graphical model based framework and system for incorporating user interactions into recommender systems. Dionysius consists of an offline model training pipeline and an online personalized recommendation system. As part of the offline training component, we learn the hidden fields for each user by considering the hierarchy of interaction signals, and replace the user profile-based fields with the learned fields, thereby not expanding the feature space at all. In other words, we make use of the existing (user, structured fields) key-value store to store these learned fields, thereby leveraging existing recommendation infrastructure designed originally to support content-based features. The intuition underlying our model is that there is typically a hierarchy in the strength of user interactions: for example, in the job recommendation domain, explicit feedback $>$ job application $>$ job view. Our model treats the user's original profile-based fields as the prior, and incorporates interactions in a hierarchical fashion, wherein the previous layer can be thought of as the prior for the next layer. The influence of the prior information decreases as the number of behavioral observations increases. In addition, the model regression coefficients are learned jointly with the user's hidden fields. The online recommendation system queries the user fields key-value store, and retrieves the hidden fields (instead of the original profile-based fields), thereby requiring very limited change to the underlying recommendation infrastructure.

While our proposed framework is general and applicable to any recommendation domain, we implemented and deployed this system for than one year as part of the job recommendation platform at LinkedIn, a professional social network with over 460M members. We validated the efficacy of our approach through extensive offline experiments with different model choices and user segments, and also using online A/B testing experiments. Our deployment of this system as part of the job recommendation engine resulted in significant improvement in the quality of retrieved results (in terms of business metrics such as VPI (views per impression) and API (applications per impression)), thereby generating improved experience and impact for millions of users.

\section{Related Work}
\label{SectionRelated}

{\noindent \bf Context-aware recommendation systems}: The notion of contextual information has been extensively studied in varied disciplines such as psychology and computer science. Bazire and Brezillon~\cite{bazire2005understanding} present and examine 150 different definitions of context from different fields, which is not surprising, given the complexity and the multifaceted nature of the concept of context. Dourish~\cite{dourish2004we} distinguishes between representational and interactional views of context. The former view considers context as static, and describes it in terms of a set of observable attributes that are known a priori. The latter view treats context as dynamic, and assumes a cyclical relationship between context and activity, where the activity gives rise to context and the context influences activity. There has been extensive work on defining and incorporating context in recommendation systems~\cite{context-handbook, 
context-www2014, 
Hariri-context-2013, 
Liu:2013:context,
Nguyen2014sigir,
context2009RecSys, 
verbert2012context,
googleRecSys2013,
zhang2014sigir}. 
There are two broad approaches to recommendation systems: content-based filtering and collaborative filtering. 
Content-based filtering~\cite{Mooney:2000, Pazzani97learningand} assumes that descriptive features of an item indicate a user's preferences. Thus, a recommender system makes a decision for a user based on the descriptive features of other items the user likes or dislikes. Usually, the system recommends items that are similar to what the user liked before. Collaborative filtering~\cite{Golbandi:2011, Herlocker:SIGIR1999, Jin:Decoupled:2003, Marlin:multiplicative:2004, collaborative:2010, Si03flexiblemixture, yu2009large} on the other hand assumes that users with similar tastes on some items may also have similar preferences on other items. 
In contrast, we propose a hierarchical graphical model to incorporate user interactions into the recommendation model. Our work can be viewed as bringing together the representational (user profile) and interactional (user interactions) views of context, in the job recommendation setting.

{\noindent \bf Job Recommendation and Professional Social Networks}: There is extensive work on how people find jobs, how they make use of their connections to obtain jobs, and how to match people and jobs, spanning diverse areas such as organizational psychology and computer science~\cite{granovetter1995getting, liu2013generating, liu2014improving, malinowski2006matching, paparrizos2011machine, wang2015user, wang2013job}. While it is desirable to take into account the person-environment (P-E) fit (subdivided further into person-organization (P-O) fit, person-vocation or occupation (P-V) fit, and person-group (P-G) or person-team fit) in addition to person-job (P-J) fit~\cite{edwards1991person,kristof1996person}, we focus primarily on the person-job fit.

\section{Problem Setting}\label{sec:problemsetting}

Before formally describing our hierarchical user interaction model, we review the basics of our existing recommender system built on logistic regression, including the modeling component and the online infrastructure.

\subsection{Logistic Regression}

Logistic regression is widely used to solve two-class classification problems with decent performance and low computational complexity, e.g., predicting the probability of a user $m$ having the observed action $y_{m,k} = 1$ on item $k$. 
The key to build a logistic regression model with good performance is to introduce valuable features. Content-based models can be implemented by introducing content-related features across structured user fields and/or structured item fields as part of the model, in which the feature space is typically static and limited. For instance, in the job recommendation application, structured user fields could include job title, seniority, skills, work experience, education experience, skills, and so on. Structured item (job) fields could include job title, function, industry, company, seniority, skills, description, and so on. User-job similarity features could be similarity between any combination of user field and item field. These features are easy to compute in both the offline training pipeline and the online prediction stage. 

We denote the feature vector between user $m$ and item $k$ as $\mathbf{x_{m,k}}$ and the coefficient vector as $\beta$. The probability could be estimated based on all available features as follows:
\begin{equation}
p(y_{m,k} = 1) = \frac{1}{1+exp\{-\mathbf{\beta^T}\mathbf{x_{m,k}}\}}\notag
\end{equation}

Assuming that the prior distribution of each model parameter is a Gaussian distribution centered on zero, the optimal coefficient vector $\beta$ can be learned from the training data using {\em maximum a posteriori probability (MAP)} estimation.

\subsection{Online Recommendation System}
\begin{figure}
\centering
\includegraphics[width=0.5\textwidth]{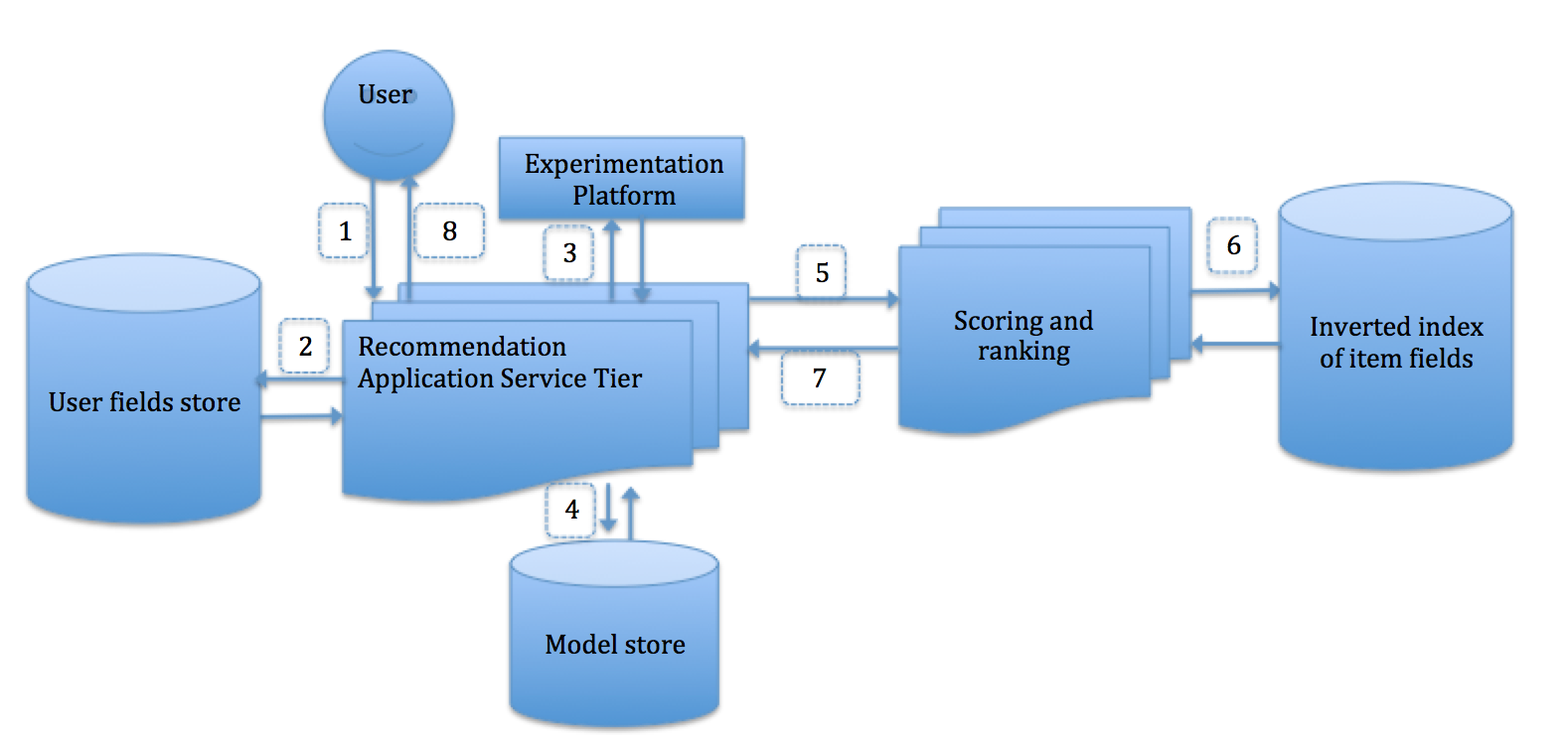}
\caption{Online recommendation system architecture at LinkedIn}
\label{figureSystem}
\end{figure}

We next briefly describe the overall design and architecture of the online recommendation system infrastructure (Figure~\ref{figureSystem}).
The following steps take place after a call to the recommendation service is made for a user (the step numbers below correspond to those shown in the diagram):
\begin{enumerate}
\item When the user visits the LinkedIn web service, a request is sent to the distributed recommendation application service tier.
\item The recommendation service retrieves the structured user fields data from a distributed key-value store.
\item It obtains the experimental treatment for the user from an external A/B testing platform service.
\item Given the machine-learned model(s) that is/are identified by the experimental treatment, the recommendation service retrieves the model(s) from the model store.
\item The recommendation service sends the request to the distributed scoring and ranking component with the target user fields and corresponding models.
\item The scoring and ranking component forms a query based on user fields, and queries against an inverted index system (such as Lucene or Galene~\cite{LinkedInGalene}) of items to retrieve a candidate set of items to score.
\item The scoring and ranking component calculates the score of candidates using a machine learned model and returns the top ranked list of items.
\end{enumerate}

\subsection{Problems and Challenges}
A fundamental problem with content-based models is that they are based primarily on the user's profile. However, two users with near identical profiles may have very different preferences for say, job recommendations: the first user may prefer jobs with similar titles as her current position from a few selected companies, while the second user may prefer jobs with identical title as her current position but from a wider range of companies.
The natural solution is to introduce dynamic user interaction signals in the logistic regression model, which simulates the idea of collaborative filtering. 
For example, if user $m$ applied to job $k$, we could incorporate this signal into the model. A common practice is to introduce user/item ID-level regression coefficients in addition to the global regression model in a Generalized Linear Model (GLM) setting, which is referred to as generalized linear mixed model (GLMix)~\cite{glmix, glmixLinkedin} in the statistics literature and recent work in industry. In our case, we can introduce the outer product between user ID $m$ and job features $j_k$, as well as between job ID $k$ and user features $u_m$. The intuition is that these ID-level features could capture job fields that are the best match for user $m$ and user fields that are the best match for job $k$. 

\begin{equation}
p(y_{m,k} = 1) = \frac{1}{1+exp\{-\mathbf{\beta^T}\mathbf{x_{m,k}} + \mathbf{\alpha_m}^Tj_k + \mathbf{\gamma_k}^Tu_{m}\}}
\notag
\end{equation}
where $\mathbf{\beta}$ is the coefficient vector of the global regression model, and $\mathbf{\alpha_m}$ and $\mathbf{\gamma_k}$ are the coefficient vectors specific to user $m$ and job $k$ respectively. $j_k$ denotes job features and $u_m$ denotes user features.

The key challenge we initially faced with such GLMix model is the high computational complexity due to the significant expansion of the feature space. Considering the scenario with 100M+ users and 1000 non-zero coefficients on job features for each user, this approach introduces more than $10^{11}$ features to learn in the model. A potential approach to address this challenge is to apply dimension reduction methods (such as feature hashing), which however reduces the ease of interpretability of the model. Another challenge we encountered for adopting and deploying GLMix model is that it involves model coefficients for every user and every job, thereby requiring a fundamental redesign of the existing online recommendation infrastructure. 
Thus, although we intuitively knew that GLMix model is likely to perform better due to the significant expansion of feature space (from about 100 to several thousand fields per user), as subsequently demonstrated by our colleagues in~\cite{glmixLinkedin}, we still had a business imperative to improve the quality of job recommendations while operating under the confines of the existing online recommendation system. Making use of various feedback / interaction signals from users was one of the best bets to achieve this goal considering that the existing system was using only user profile features; however, we could not afford to wait until the online recommendation and ranking infrastructure was redesigned, implemented, and sufficiently stable for our use case.

Thus, we addressed the following problem: {\em How do we incorporate user item interaction signals as part of the relevance model in a large-scale personalized recommendation system such that, (1) the ability to interpret the model and explain the recommendations is retained, and (2) the existing infrastructure designed for the (user profile) content-based model can be leveraged without significant change?}

\section{Dionysius: A Framework for Modeling Hierarchical User Interactions}
We next present the desiderata for addressing our problem and the notations/preliminaries needed to describe our framework. Then, we describe our hierarchical user interaction model, followed by a brief overview of how we implement and deploy as part of the large-scale recommendation system.

\subsection{Desirable Properties}
We first highlight the desirable properties of a framework for modeling user item interactions in our recommendation setting:\\

{\noindent \em Interpretability and explainability of recommendations}: The model should be easy to interpret, and the recommendations are easy to explain.\\

{\noindent \em Differential weights for different action types}: Stronger interaction types (such as applications) should be given greater weight than weaker ones (such as views).\\

{\noindent \em Graceful fallback}: The model should be able to handle users with significant interaction activity as well as users with no interaction activity. In other words, it should gracefully fallback to the user profile-based setting when there is no interaction activity.\\

{\noindent \em Infrastructure and deployment considerations}: The proposed approach should not require significant change to the existing recommendation infrastructure, which is designed as a content based recommendation system.

\subsection{Preliminaries}
\label{SectionProblem}
For concreteness, we describe our approach in terms of the job recommendation application, where the goal is to predict the probability for a user $m$ applying to a job $k$. With that information, the system can rank all potential jobs according to the probability and recommend top ones to user $m$.
We use the following notations. 

\begin{itemize}
\item
$m = 1, 2, ..., M$: the index of the user.
\item
$k = 1, 2, ..., N$: the index of the job.
\item
$y_{v,m,k}$: the binary label that indicates the user's viewing behavior for job $k$. If $y_{v,m,k}=1$, it indicates that the user $m$ clicks to view job $k$. Otherwise $y_{v,m,k} = -1$.
\item
$y_{a,m,k}$: the binary label that indicates the user's application behavior for job $k$. If $y_{a,m,k}=1$, it indicates that the user $m$ clicks to apply to job $k$.  Otherwise $y_{a,m,k} = -1$.
\item
$\mathbf{u_{p,m}}$: the vector of profile-based fields associated with user $m$. The vector includes static demographic features that are derived from the user profile information.
\item
$\mathbf{u_{v,m}}$: the vector of view-based fields associated with user $m$. The vector is learned from the data according to the user's viewing behavior.
\item
$\mathbf{u_{a,m}}$: the vector of application-based fields associated with user $m$. The vector is learned from the data according to the user's application behavior.
\item
$\mathbf{j_k}$: the vector of fields associated with job $k$. The vector includes static features that are derived from the job description.
\item
$\mathbf{x_{m,k}}$: the feature vector that is associated with the user $m$ profile and job $k$. It might include user profile-based fields $\mathbf{u_{p,m}}$, job profile-based fields $\mathbf{j_k}$, and similarity-based features between user profile-based fields $\mathbf{u_{p,m}}$ and job profile-based fields $\mathbf{j_k}$.
\item 
$D = \{D_1, ..., D_m, ..., D_M\}$: The observed data of all users.
\item
$D_m = \{ y_{v,m,k}, y_{a,m,k}, \mathbf{u_{p,m}}, \mathbf{j_k}\}$: A set of observed data associated with user $m$. 
Each observation is associated with four parts: the user viewing behavior, the user application behavior, the user profile-based fields, and the job fields .
\item
$\beta_v$: the $d$-dimensional vector of regression coefficients to predict the user's viewing behavior $y_{v,m,k}$.
\item
$\beta_a$: the $d$-dimensional vector of regression coefficients to predict the user's application behavior $y_{a,m,k}$.
\item
$\sigma_v$: variance of the user view-based feature vector $\mathbf{u_{v,m}}$.
\item
$\sigma_a$: variance of the user application-based feature vector $\mathbf{u_{a,m}}$.
 
\end{itemize}

\subsection{Hierarchical User Interaction Model}\label{sec:hiermodel}

We next describe the hierarchical graphical model to incorporate user interactions into item recommendations (presented in terms of job recommendation application, for concreteness). We learn the user hidden fields vector by considering their interaction signals (job views, applications). In the global regression model, we replace the user profile-based vector with the user hidden fields vector, thereby not expanding the feature space at all. 
The user hidden fields vector is learned based on their job viewing and applying behavior as well as other explicit positive/negative feedbacks on the recommended jobs. The intuition underlying our model is that there is a hierarchy in the strength of user activity: job application > job view. Note that we describe our model with two types of user behaviors while it is straight-forward to add other types of interactions (such as saving a job) in the model as an additional layer in the hierarchy. We describe components of the model as follows.
\begin{description}
\item[Model parameters]. The first component specifies the parameter likelihood in the model. In particular, $\beta_v$ is the coefficient vector for the regression model which predicts the user viewing behavior. $\beta_a$ is the coefficient vector for the regression model which predicts the user applying behavior. $\sigma_v$ controls the variance of the distribution where user view-based fields $u_{v,m}$ is drawn from. $\sigma_a$ controls the variance of the distribution where user application-based fields $u_{a,m}$ is drawn from.

We assume that parameters are sampled from Gaussian distributions and Inverse-Gamma distributions respectively: 
$\beta_v \sim N (\mu_{\beta_v}, \sigma_{\beta_v})$;
$\beta_a \sim N (\mu_{\beta_a}, \sigma_{\beta_a})$;
$\sigma_v \sim \Gamma^{-1} (\alpha_{\sigma_v}, \gamma_{\sigma_v})$;
$\sigma_a \sim \Gamma^{-1} (\alpha_{\sigma_a}, \gamma_{\sigma_a})$.

For easier interpretation, we denote \\$\phi = (\mu_{\beta_v}, \sigma_{\beta_v}, \mu_{\beta_a}, \sigma_{\beta_a}, \alpha_{\sigma_v}, \gamma_{\sigma_v}, \alpha_{\sigma_a}, \gamma_{\sigma_a})$

\item[User fields vector]. The second component specifies the likelihood of the user fields vector. The user profile-based fields vector, $u_{p,m}$ is first computed from the user's profile. We assume that (1) the user view-based fields vector, $u_{v,m}$ follows Gaussian distribution with $u_{p,m}$ as mean and $\sigma_v$ as variance, and (2) the user application-based fields vector, $u_{a,m}$ follows the Gaussian distribution with $u_{v,m}$ as mean and $\sigma_a$ as variance.
\begin{align}
\mathbf{u_{v,m}} \sim N (\mathbf{u_{p,m}}, \mathbf{\sigma_{v}})\\\notag
\mathbf{u_{a,m}} \sim N (\mathbf{u_{v,m}}, \mathbf{\sigma_{a}})\\\notag
\end{align}
In practice, $\sigma_v$ and $\sigma_a$ could be tuned manually to control the weight of the prior fields that come from the user profile. As the variance $\sigma_v$ becomes larger in magnitude (component-wise), the model gives greater weight to the user viewing behavior and cares less about the user's profile. Similarly, as the variance $\sigma_a$ becomes larger in magnitude (component-wise), the model gives greater weight to the user application behavior and cares less about the user's profile / viewing behavior.

\item[User interaction signal]. The third component specifies the likelihood of the user interaction signal. In the former one, the user viewing behavior $y_{v,m,k}$ is dependent on the user view-based vector $u_{v,m}$ and the regression model $\beta_v$. In the latter one, the user application behavior $y_{a,m,k}$ is dependent on the user application based vector $u_{a,m}$ and the regression model $\beta_a$. We use logistic regression as the core model to predict user action. In the following equation, $f$ function calculates the feature given the job fields and user fields. It could be cosine similarity, jaccard similarity and so on.
\begin{align}
p(y_{v,m,k}|\mathbf{u_{v,m}}, \beta_v) = \frac{1}{1+exp(-y_{v,m,k} ({\beta_v}^T f(j_k, \mathbf{u_{v,m}} )))} \\\notag
p(y_{a,m,k}|\mathbf{u_{a,m}}, \beta_a) = \frac{1}{1+exp(-y_{a,m,k} ({\beta_a}^T f(j_k, \mathbf{u_{a,m}} )))} \notag
\end{align}
\end{description}

\begin{figure}
\centering
\includegraphics[height=0.35\textwidth]{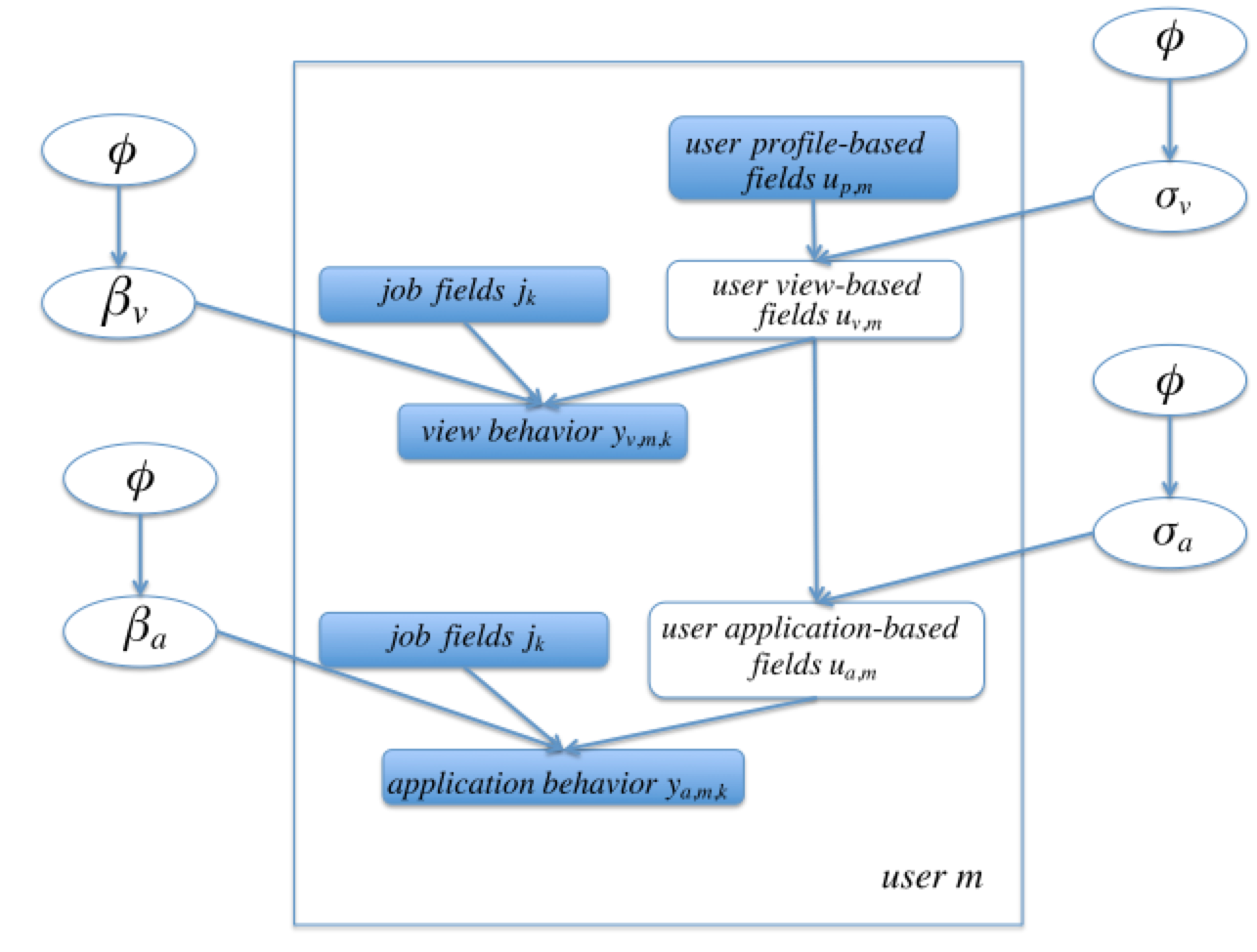}
\caption{Illustration of dependencies of variables in the hierarchical user interaction model. It shows the observed dataset of user $m$, i.e., $D_m = \{ y_{v,m,k}, y_{a,m,k}, \mathbf{u_{p,m}}, \mathbf{j_k}\}$. All variables in shades are observed ones while others are hidden variables in the model. As shown in the plot, the user view-based fields $u_{v,m}$ is drawn from distribution of user profile-based fields $u_{p,m}$ and variance $\sigma_v$. Similarly, the user application-based fields $u_{a,m}$ is drawn from distribution of user view-based fields $u_{v,m}$ and variance $\sigma_a$. The user view behavior $y_{v,m,k}$ is dependent on the regression model $\beta_v$, the job fields $j_k$ and the user view-based fields $u_{v,m}$. Similarly, the user application behavior $y_{v,m,k}$ is dependent on the regression model $\beta_a$, the job fields $j_k$ and the user application-based fields $u_{a,m}$.}
\label{figureModel}
\end{figure}

Assuming that the data is independently and identically distributed, we can present the data likelihood as:

\begin{align}
p(D|\phi) &= \int p(D, \theta_g|\phi) d \theta_g\\\notag
&= \int p(D| \theta_g, \phi) p(\theta_g|\phi) d \theta_g \\\notag 
&= \int [\prod_{m=1}^M p(y_m| \theta_g,\phi)] p(\theta_g|\phi) d\theta_g \\\notag 
\end{align}

where $\theta_g = (\beta_a, \beta_v, \sigma_a, \sigma_v)$ is a random variable denoting the joint distribution of the global random variables.

In particular, the data likelihood for user $m$ can be presented as 
\begin{align}
&p(y_{m}| \theta_g, \phi)  
= \int p(y_{m}|\theta_m, \theta_g, \phi) p(\theta_m|\theta_g, \phi) d \theta_m \\\notag 
&p(y_{m}|\theta_m, \theta_g, \phi)p(\theta_m|\theta_g,\phi) \\\notag
= &[\prod_{k=1}^{K_m} [p(y_{v,m,k}|\mathbf{u_{v,m}}, \beta_v)p(y_{a,m,k}|\mathbf{u_{a,m}}, \beta_a)] ]\\\notag
&p(u_{a,m}|u_{v,m}, \sigma_a) p(u_{v,m}|\sigma_v) \\\notag
\end{align}

where $\theta_m = (\mathbf{u_{v,m}}, \mathbf{u_{a,m}})$ is a random variable denoting the joint distribution of the view-based vector and application-based vector random variables for each user $m$.

Maximizing the likelihood of $p(D | \phi)$ is equivalent to maximizing the log likelihood, $L(D | \phi) = \ln p(D | \phi).$

There is no known closed-form solution for the estimation of model parameters. We follow the Bayesian method~\cite{VarBayesian} to derive an Expectation-Maximization (EM) based iterative process to find an approximate solution. In the E step, we fix the regression model and tune the user interaction-based vector. In the M step, we fix the user interaction-based vector and learn the regression model accordingly. The iterative process converges when there is minimal change in regression model coefficients and user interaction-based vectors.

Thus, our proposed approach can be summarized as follows:
\begin{itemize}
\item
We propose a hierarchical graphical model to incorporate user interactions into the recommendation model. 
\item
The user hidden fields vector is learned based on the user's job viewing behavior and job applying behavior.
\item
Three layers of user information would be leveraged hierarchically in the recommendation stage, including the user's profile, user's viewing behavior, and the user's applying behavior. The previous layer would act as the prior for the next layer. The influence of the prior information decreases as the number of behavior observations increases.
\item
The regression coefficients are learned jointly with the user's hidden fields vector.
\end{itemize}

{\noindent \bf Offline System for Generating and Incorporating User Interaction Model}: Note that there is no change to the online infrastructure, which is a key motivation for our modeling framework. 
In the offline system, the interaction based fields vectors for users are generated and pushed to the structured user fields store (a distributed key-value store) through the following steps:
\begin{enumerate}
\item Train the hierarchical user interaction model using the user's interactions and store the regression coefficients for prediction.
\item Use the trained user interaction model and the user's interactions in the last N days to predict the user's interaction-based hidden fields vector (this is updated periodically).
\item Push the user's interaction-based hidden fields vector to the distributed structured user fields store so that it can be retrieved by the online system.
\end{enumerate}

 \section{Offline Experiments}
\label{sec:exp}

We next present an extensive evaluation of our framework applied to the job recommendation application at LinkedIn. We consider the following types of log events: {\em job impression} (whether a job was shown to a user), {\em job view} (whether the user clicked on the job to view its detailed description), and {\em job application} (whether the user then applied for the job). Of these, we treat job views and job applications as the user interaction signal types. We address the following research questions through our experiments:

\begin{itemize}
\item Is it helpful to consider signals from user interactions in the model? How much lift do these signals have in terms of the overall recommendation performance?
\item Is it helpful to incorporate different types of user interaction signals in a hierarchical manner?
\item How does the performance compare for different user segments? 
\end{itemize}

\subsection{Evaluation Metrics}\label{sec:metrics}

In the offline evaluation, we compute the area under receiver operating characteristic curve (ROC AUC), which represents the quality of the recommendation system (viewed as a binary classifier). A (user, job) pair is labeled as a positive example if the user applies to the job during the evaluation period, and as a negative example otherwise. The ROC curve is obtained by plotting the true positive rate (recall) against the false positive rate at various choices of the threshold in the recommendation model. ROC AUC can be interpreted as the probability that the binary classifier will rank a randomly chosen positive instance higher than a randomly chosen negative instance.

\subsection{Models to Compare}
\label{SectionModels}
\begin{table}
\caption{Models to compare}
\centering{
\begin{tabular}{c||cc} \hline
Model&User interaction signal&hierarchy or not\\\hline
M-baseline&None&None\\
\hline
M-view&job views&None\\
\hline
M-apply&job applications&None\\
\hline
M-viewApply&job views, job applications&Hierarchical model\\
\hline
\end{tabular}
}
\label{TableModel}
\end{table}

We compare the basic regression model, as well as the user interaction models with different components. As shown in Table~\ref{TableModel}, \textbf{M-baseline} is the basic regression model with no user interaction signals. \textbf{M-view} is the user interaction model with only viewing behavior. \textbf{M-apply} is the user interaction model with only application behavior. \textbf{M-viewApply} is the hierarchical user interaction model with both viewing behavior and application behavior.

All models use the same set of features for the core regression model. Features correspond to similarity between different combinations of structured user fields (such as job title, function, seniority, skills, work experience, education experience, ...) and structured job fields (such as job title, function, seniority, skills, description, ...).

Note that since our goal is to leverage the existing recommendation system infrastructure, we limit our comparisons to models described above. 
In particular, we do not compare against more complex models such as those based on collaborative filtering or matrix factorization since these require significant infrastructure change, and hence are not feasible candidates for deployment in our setting.

\subsection{Dataset}
\label{SectionDataset}

\subsubsection{Training Stage}

\begin{figure}
\centering
\includegraphics[width=0.45\textwidth]{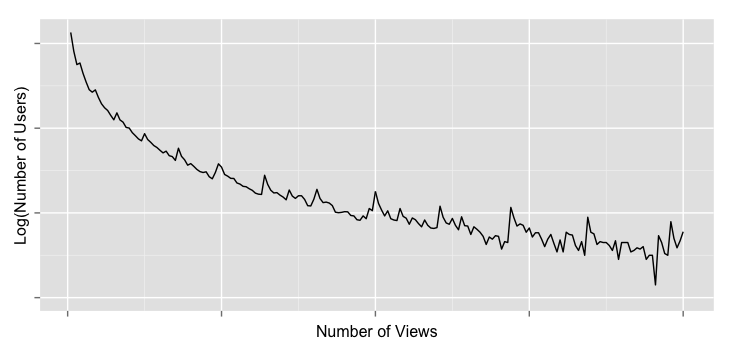}
\includegraphics[width=0.45\textwidth]{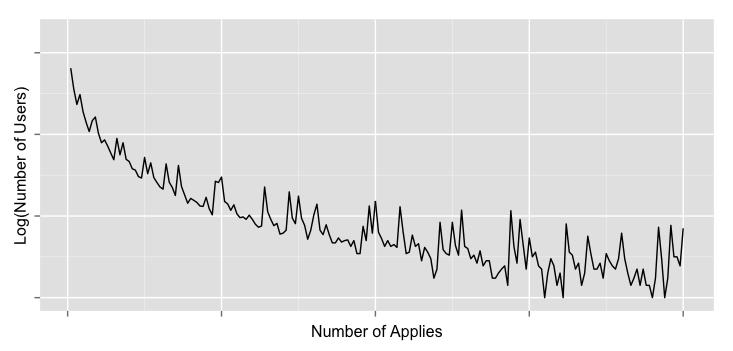}
\includegraphics[width=0.45\textwidth]{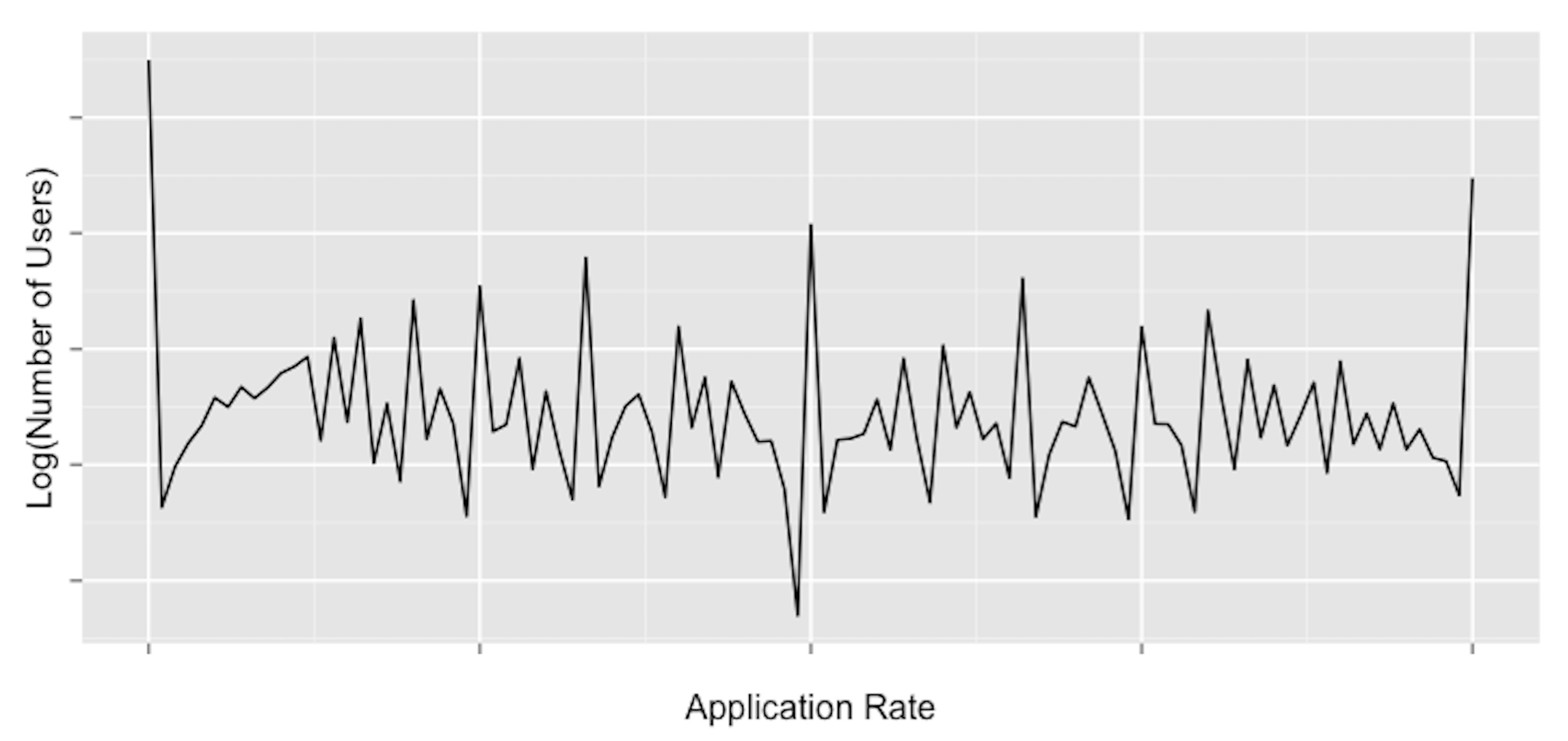}
\caption{Distribution of user interaction in the dataset. In all three plots, Y-axis corresponds to log(number of users). X-axis corresponds to the number of views, the number of applies, and the job application per view rate (APV) respectively. Exact number is omitted due to business reasons.}
\label{figureData}
\end{figure}

In the model training stage, we used two different datasets, for learning user interaction-based fields and for training the logistic regression model respectively. 
\begin{description}
\item[Training: dataset for tuning user interaction-based fields] We first collect the data to learn user interaction-based fields if a user viewed a job or applied to a job from 2016/01/01 to 2016/01/09. This sample data contains millions of users that applied to a large number of jobs in that period\footnote{Exact number is omitted due to business reasons.}. Figure~\ref{figureData} shows distributions of user interactions. 
We observe that a few users applied to or viewed a lot of jobs while the majority of users applied to or viewed only a few jobs. 
Figure~\ref{figureData} also shows the distribution of the application per view (APV) rate (that is, the ratio of the number of applications to the number of views). There are two peaks in the plot, corresponding to users that did not apply to any jobs they viewed and to users that applied to almost all jobs they viewed respectively. Consequently, we evaluate the performance of different models with respect to each of these two user segments in \S\ref{sec:activepassive}.

\item[Training: dataset for training logistic regression model] We then collect the dataset to learn the logistic regression model with user interaction-based fields for the time period from 2016/01/10 to 2016/01/15. If user $m$ applied to job $k$, we assign a positive label to $(u, k)$ pair. If user $m$ viewed or applied to job at position $t$, we treat the jobs shown up to position $t-1$ as having negative labels for $m$. In addition, we generate random negative data by randomly sampling from the set of all jobs that have been applied on a particular day. If user $m$ applied to jobs on some day, we first obtained all jobs that have been applied by other users on that day and removed jobs that were applied by user $m$. We then randomly sampled from this set, and assigned negative labels to these jobs for user $m$.
\end{description}

\begin{table*}
\caption{Offline ROC AUC for different models and user segments. Users are divided into five segments: \textit{high App, high APV; high App, low APV; zero App, high View; zero App, low View; zero App, zero View}, as described in \S\ref{SectionDataset}.}
\centering
\scalebox{0.8}{
\begin{tabular}{|c||c|ccccc|} \hline
&all users&high App, high APV&high App, low APV&zero App, high View&zero App, low View&zero App, zero View\\\hline
M-baseline&0.612&0.612&0.571&0.602&0.604&0.618\\\hline
M-view&0.638 (+4.2\%)&0.663&0.604&0.606&0.619&0.669\\\hline
M-apply&0.643 (+5.1\%)&0.673&0.603&0.603&0.605&0.675\\\hline
M-viewApply&\textbf{0.644 (+5.2\%)}&\textbf{0.682}&\textbf{0.606}&\textbf{0.612}&\textbf{0.623}&\textbf{0.677}\\\hline
\hline
\end{tabular}
}
\label{TableRecOffline}
\end{table*}

\begin{table*}
\caption{Example of user profile, along with the user's activities and recommendations from each model. We show only the job title due to privacy reasons. Note that jobs with the same title might refer to different unique jobs.}
\centering
\scalebox{1.0}{
\begin{tabular}{|c||c|} \hline
User Profile Title&Business Analyst\\\hline
Jobs viewed by the user&Software Engineer, Product Manager, Business Analyst
\\\hline
Jobs applied to by the user&Product Manager, Business Analyst\\\hline
\hline
\end{tabular}
}
\scalebox{1.0}{
\begin{tabular}{|c||c|} \hline
\textbf{M-baseline}&
Business Analyst, Consultant - Business Intelligence and Business Analytics, 
Data Analyst
\\\hline
\textbf{M-view}&Business Analyst, Product Manager, Software Engineer\\\hline
\textbf{M-apply}&Business Analyst, Product Manager, Product Manager\\\hline
\textbf{M-viewApply}&Business Analyst, Product Manager, Software Engineer\\\hline
\hline
\end{tabular}
}
\label{TableExample}
\end{table*}

\begin{table*}
\caption{Online A/B testing performance comparison. Relative improvements are shown for each business metric. \textbf{M-viewApply} performs significantly better than \textbf{M-baseline} with respect to key metrics, API (job application per impression) and VPI (job view per impression).}
\centering
\scalebox{1.0}{
\begin{tabular}{|c||ccc|cc|} \hline
Model&\# of impressions&\# of applications&\# of views&{\bf API}&{\bf VPI}\\\hline
M-viewApply vs. M-baseline&+4.1\%&+7.8\%&+7.7\%&{\bf +3.6\%}&{\bf +3.5\%}\\
\hline
\end{tabular}
}
\label{TableRecOnline}
\end{table*}

\subsubsection{Testing Stage}

In the model testing stage, we used two datasets, for learning user interaction fields and for providing recommendations respectively. 

\begin{description}

\item[Testing: dataset for learning user interaction fields] We first collect user interactions from 2016/01/16 to 2016/01/25 to learn the user interaction-based fields. We segment users into five different groups based on their interactions in this time period: \textit{high App, high APV; high App, low APV; zero App, high View; zero App, low View; zero App, zero View}. \textit{high App} corresponds to users that have applied to at least one job while \textit{zero App} corresponds to users that have not applied to any job. We consider the distribution of users with respect to their job application per view rate (APV), and categorize users at $75^{th}$ percentile and above as corresponding to \textit{high APV} and users below $75^{th}$ percentile as corresponding to \textit{low APV}.
Similarly, \textit{high View} corresponds to users with higher number of job views, \textit{low View} corresponds to users with lower number of job views, and \textit{zero View} corresponds to users that did not view any job's detailed description.

\item[Testing: dataset for providing recommendations] We collect user application data from 2016/01/26 to 2016/01/30. The positive data and negative data are labeled similarly as we did in the training stage. All user interaction-based fields that were learned in the previous step are used here in the recommendation stage, along with the regression model that was learned in the training stage. Note that if the user does not have any interaction in the previous step, user interaction-based fields would gracefully fall back to the user's profile-based fields.
\end{description}

\subsection{Performance Analysis: User interaction models vs Baseline model}\label{sec:offlineevaluation}

We present the performance of different models in Table~\ref{TableRecOffline}.
We first validate the underlying premise of user interaction models by evaluating whether incorporating user interaction signals leads to better performance. Indeed, we observe that both \textbf{M-view} and \textbf{M-apply} lead to significant improvement (4.2\% and 5.1\% respectively) in ROC AUC compared to the baseline model with no interactions, \textbf{M-baseline}. Further, \textbf{M-apply} performs better than \textbf{M-view}, validating that applying to a job is a stronger signal than merely viewing a job's detailed description, and that a user's past job application behavior contains more reliable signals for predicting jobs that the user is likely to apply in future.

Table~\ref{TableExample} provides an illustration of a representative, real user's profile, the user's job related interactions, and the recommendations from each model. This user currently works as a \textit{Business Analyst} and has viewed jobs with the following titles: \textit{Software Engineer, Product Manager, Business Analyst}. Note that while the user viewed some jobs with the same title as the user's current job title (stated in the user profile), the user also viewed jobs with other titles. From amongst the viewed jobs, this user chose to apply to jobs with titles: \textit{Product Manager, Business Analyst}. Based on this information, the baseline model, \textbf{M-baseline} recommends jobs with titles similar to the user's current title (\textit{Business Analyst}), since it is not aware of the user's intention to explore \textit{Software Engineer} / \textit{Product Manager} jobs. However, the user interaction models successfully incorporated such interaction signals into the recommendations. \textbf{M-view} recommends jobs that are consistent with the user's viewing behavior while \textbf{M-apply} recommends jobs that are consistent with the user's application behavior. The hierarchical model \textbf{M-viewApply} considers signals from both types of interactions and recommends jobs accordingly. Note that the actual job recommendations from \textbf{M-view} and \textbf{M-viewApply} are different although they share the same job title. Additional job details are not shown due to privacy reasons.

\section{Online Experiments}

\subsection{Online A/B Testing Setup}
Next, we evaluate the performance of different models by deploying to production job recommendation systems serving live traffic of users at LinkedIn. We compare the best performing user-interaction model, \textbf{M-viewApply} from our offline experiments against the baseline model, \textbf{M-baseline}.
For the online experiment, we randomly select 5\% users that visit the site for each of these two models and present the corresponding job recommendations to each user group. We report the relative difference in performance between two user buckets.
A significance level of 0.05 with the $t$-test is used to compare two models. We first let each model serve the live traffic for one week to account for the novelty effect (active job seekers tend to apply to any new jobs they see due to a change in the recommendation model, and hence initial results could be attributed to such novelty effect in addition to the underlying model improvements) and report the comparison of the performance of the models in the subsequent time period.

We use the following two business metrics for online evaluation: (1) the job view rate (VPI), which is defined as the ratio of the number of jobs viewed by users to the number of job impressions (number of jobs presented to users); (2) the job application rate (API), which is defined as the ratio of the number of jobs applied to by users to the number of job impressions. Higher job view and application rates correspond to more relevant jobs being shown to users, and hence are desirable. We also present the number of impressions, the number of views, and the number of applications. For all of these metrics, we report only the relative change between two models, and omit the absolute numbers due to business reasons.

\subsection{Performance Analysis}
We observe significant performance improvements in the online A/B testing experiment, agreeing with our intuition. These results, presented in Table~\ref{TableRecOnline}, demonstrate that the proposed hierarchical model, \textbf{M-viewApply}, which incorporates user interaction signals, performs significantly better than the baseline. In particular, the {\bf job view rate (VPI)} and the {\bf job application rate (API)} improve by as much as {\bf 3.5\%} and {\bf 3.6\%} respectively. These results validate that whenever present, a user's job view and application behavior inputs are very useful signals for providing relevant job recommendations since a user's job related activity may deviate from what is stated in the user's profile. Further, from these results, we can conclude that our proposed model is capable of incorporating such user interaction signals to provide more relevant recommendations, while gracefully falling back to the information present in the user profile when there are no interaction signals.

\section{Lessons Learned from Practice}\label{sec:lessons}
We next highlight a few lessons learned during experimentation with real datasets and over the course of the deployment of our system in production at LinkedIn.

\subsection{Performance analysis for different user segments}\label{sec:activepassive}

When we compared the performance over all the users, we observed that \textbf{M-viewApply} performs significantly better than \textbf{M-baseline} (Table~\ref{TableRecOffline}). However, in order to truly understand each model's prediction power, we decided to analyze their performance on the five user segments, \textit{high App, high APV; high App, low APV; zero App, high View; zero App, low View; zero App, zero View}.

The first segment, \textit{high App high APV} corresponds to users with many job views and comparably many job applications. All user interaction models perform significantly better than the baseline model for this segment. In addition, \textbf{M-viewApply} performs better than \textbf{M-apply} which is better than \textbf{M-view}, thereby demonstrating the benefit of using a hierarchical model.

The second segment, \textit{high App low APV} corresponds to users with many job views but relatively lower number of job applications. In this case, \textbf{M-view} performs slightly better than \textbf{M-apply} since the former incorporates more signals from the user viewing history. Interestingly, we observe that \textbf{M-viewApply} performs slightly better than both \textbf{M-view} and \textbf{M-apply}.

The third and fourth segments correspond to users with at least one job view but no job applications. On the one hand, \textbf{M-apply} and \textbf{M-baseline} have roughly the same performance since no application signal is available in these segments. On the other hand, \textbf{M-view} performs slightly better than \textbf{M-baseline} while \textbf{M-viewApply} performs significantly better. An intuitive explanation is that \textbf{M-viewApply} is able to discriminate amongst the viewed jobs to better predict jobs that a user is likely to apply to, since this model is provided signals from both job views and job applications across users.

The last segment, \textit{zero App zero View} corresponds to users with no interactions in the prediction stage. Surprisingly, we observe significantly better performance from all user interaction models, compared with the baseline model, \textbf{M-baseline}. A possible explanation is that the logistic regression model learned (over training data from all users) in the three user interaction model settings has better prediction power than that for the baseline model, which is trained without considering user interaction signals.

\subsection{From deployment to gradual retirement}
Our system was deployed in production at LinkedIn for nearly 18 months, from early 2015 to late 2016. During this period, a new ranking platform based on LASER~\cite{Agarwal:2014:LSR:2556195.2556252} and Galene~\cite{LinkedInGalene} was developed at LinkedIn, and gradually tested and deployed in production to replace the legacy infrastructure which served several dozens of client applications. The job recommendation application was the first to migrate to this new platform, which subsequently enabled the adoption of the GLMix model with richer features as well as easier feature modeling efforts. The full deployment of the GLMix model (that is, to 100\% of users for all job recommendation application scenarios, including presenting job recommendations as part of LinkedIn Jobs homepage, personalized feed, job related emails sent to users, and job recommendations shown on LinkedIn company pages) using the new platform was performed over the course of several months, after which we have been able to retire the {\em Dionysius} system. During this interim period, Dionysius as well as the new ranking platform with the new GLMix model coexisted in production; different random subsets of users were served by either of these systems, depending on the experimental treatment bucket that the users belonged to. We performed such migration in an incremental fashion, where we progressively moved more users from Dionysius to the new platform typically on a weekly basis. We carefully monitored the load on our systems (especially on Mondays when we received larger traffic), performed extensive tests comparing the quality of the results from both systems, and gradually shifted the user traffic entirely to the new platform, prior to retiring Dionysius system.
Although our system is no longer deployed in production, we still believe that our framework and practical experience described in this paper can potentially benefit data mining researchers and practitioners who are faced with the similar challenge of incorporating user feedback signals into an existing large-scale personalized recommendation platform, either due to the need to use the existing infrastructure, or due to the desire to have simple explainable models.

\section{Conclusion and Future Work}
\label{sec:conclusion}
We proposed {\em Dionysius}, a hierarchical graphical model based framework for incorporating user interactions into recommender systems. Our framework enables the incorporation of user item interaction signals as part of the relevance model in a large-scale personalized recommendation system, with {\em minimal change to the existing recommendation infrastructure}, while retaining the ability to interpret the model and explain the recommendations. As part of our proposed model, we learned a hidden fields vector for each user by considering the hierarchy of interaction signals, and replaced the user profile based vector with this learned vector, thereby not expanding the feature space at all. Our implementation and deployment of this system as part of the job recommendation platform at LinkedIn demonstrated the efficacy and practicality of our framework, and also resulted in significant improvement in the quality of the recommendation results for millions of users.

A promising direction is to extend and apply this framework for items, that is, incorporate signals from users that interacted with the item (e.g., viewed/applied to a job) to enhance the representation of the item in a hierarchical fashion, wherein the strength of the interaction is factored in. Another fruitful direction to pursue is to apply our framework for associations beyond interactions, by considering relationship in the structured fields. For example, in the job recommendation application, we can take into account jobs that have the same title as the user, jobs that have the same title and skills as the user, and so on, and apply the hierarchical model to enhance the representation for users with no interaction activity (including new users). Likewise, we can apply such association to enhance the representation for new jobs or those with no views or applications from users.

\bibliographystyle{abbrv}
\bibliography{paper}

\end{document}